\documentclass[fleqn,10pt]{wlscirep}
\usepackage[utf8]{inputenc}
\usepackage[T1]{fontenc}
\usepackage{siunitx}
\usepackage{textcomp}
\usepackage{gensymb}
\usepackage{float}
\usepackage{xcolor}
\usepackage[normalem]{ulem}
\title{Controlling the Size of Nanoparticles Using a Magnetic Field: A Sphere Packing Approach}

\author[1]{Yazeed Tawalbeh}
\author[1]{Marwa Ghazi}
\author[1,2,*]{Mauro Fernandes Pereira}
\affil[1]{Department of Physics, Khalifa University of Science and Technology, 127788 Abu Dhabi, United Arab Emirates}
\affil[2]{Institute of Physics, Czech Academy of Sciences, 18221 Prague, Czech Republic}

\affil[*]{mauro.pereira@ku.ac.ae}


\begin{abstract}
We present an analytical framework that predicts and controls nanoparticle size through external magnetic fields, uniting first-principles thermodynamics with a sphere packing approach.  Calibrated to diamagnetic silver nanoparticles (20 nm at zero field and 5 nm at 250 mT), the model yields a closed-form relation between radius and field that reproduces the observed shift in most-probable size. Within the limits of classical capillarity and spherical demagnetization, the field lowers the nucleation barrier and drives the distribution toward smaller particles. {Our results are robust for radii above $\approx3$ nm ($\approx5740$ atoms). Below this scale non-extensive effects likely dominate, as discussed in detail in Supplementary Information}. The approach generalizes to both diamagnetic and paramagnetic systems and the limitations expected for very small or ferromagnetically ordered nanoparticles are discussed.
\end{abstract}
\begin{document}

\maketitle

\thispagestyle{empty}

\noindent 
\section*{Introduction}
 {
Nanoparticles (NPs) exhibit unique chemical, optical and physical properties compared to their bulk counterparts, mainly due to quantum confinement effects and their high surface area–to–volume ratio. These properties make NPs central to a wide range of applications. 
In biomedicine,  Mn-substituted magnetite superparamagnetic nanoparticles have shown promise for cancer treatment \cite{10.1016/j.jallcom.2023.172999}, while manganese ferrite (MnFe2O4) magnetic NPs are being explored for targeted drug delivery and magnetic hyperthermia therapy \cite{10.1016/j.matpr.2023.05.456}.
In electronics, NPs of Mn-substituted Y3Fe5-xMnxO12 have potential for use in microwave, spintronics and other advanced electronic devices \cite{doi.org/10.1016/j.ceramint.2024.06.067}.
In environmental applications,  nano-silica clay and acid-treated earthenware clays have proven effective in wastewater treatment, particularly in removing dye pollutants from aquatic systems \cite{10.1007/s11051-024-06119-8,10.1016/j.totert.2023.100038}.
In the energy sector,  synthetic biology and metabolic engineering are advancing next-generation sustainable biofuels \cite{10.1039/D5YA00118H}, with NP catalysts playing a crucial role in enhancing yields, stimulating renewable processes, and improving energy efficiency \cite{D1MA00538C}.
In photonics and biomedicine,  NPs are increasingly being employed in THz imaging and sensing technologies, as well as in innovations such as nanoparticle-assisted tissue soldering \cite{D4MA90092H, Fabrizio12, henini2011handbook, Dong:20}.
In optics and sensing,  aggregates of metallic NPs (e.g., silver or gold) make extremely efficient substrates for surface-enhanced Raman scattering (SERS) \cite{Fabrizio12}. In quantum metrology and information science, quantum squeezing plays a pivotal role in enhancing measurement precision \cite{11142662} and NPs serve as essential technology enablers, acting as efficient quantum emitters \cite{Uppu2021}.
}

A challenge in nanotechnology is to achieve a refined size and shape of NPs \cite{Ganguly2024}.  Reducing NP size and achieving a uniform ensemble of NPs are crucial for different applications such as their use as intravascularly injectable NPs (nanovectors or nanocarriers), which is of extreme relevance for the treatment of diseases, notably cancer, since photothermal therapies to destroy cancer cells require an accurate determination of the light conversion capabilities of plasmonic nanoparticles to achieve the necessary temperature-induced effects in biological tissue \cite{Naccache2017}.  In this paper, we present a theoretical framework for predicting how the size of a NP reacts to exposure of the growth medium to a magnetic field by further developing the classical nucleation theory, consistently including the effect of an external magnetic field, combined with a sphere packing correction.

{While atomistic methods such as classical density functional theory (cDFT) and electronic density functional theory (DFT) can, in principle, capture correlations and surface effects, beyond the capillarity approximation, their practical application to nanoparticle nucleation remains computationally demanding. Even the smallest nanoparticles considered here (radius $\approx$ 2 nm) contain over $10^3$ atoms, beyond the range typically tractable by DFT. Therefore, our analytical approach complements these atomistic methods by providing closed-form relations that preserve physical transparency while remaining quantitatively consistent with experimental trends. Recent reviews highlight the ongoing efforts to bridge these scales through hybrid and multiscale frameworks \cite{ Giovannini2023-fs}. More details and further references to the literature are given in ‘Results and Discussion’, under ‘Limitations and Range of Validity.}

Before we move on, an important remark is needed. The sphere-packing problem has always fascinated physicists and mathematicians, and along the centuries advanced mathematical methods have been developed to pack spheres together in a way that minimizes the empty space between them.  Historically, the first solutions for free packing came from Johannes Kepler in 1611. The formal proof for the Kepler conjecture: No packing of congruent balls in Euclidean three space has density greater than that of the face-centered cubic packing was derived by Thomas C. Hales in 2005 \cite{Hales2005}.
Even though higher dimensions are not of relevance for our work here, it is worth noticing that this is still a problem of high scientific interest, to the point that Maryna Viazovska won the Fields Medal in 2022 for her work on 8 and 24 dimensions. \cite{Vyazovska8,Vyazovska24}. Research in other dimensions continues.
In this paper, the 3-D packing of small identical spheres in a larger sphere is used. We have found analytical approximations for the open-source numerical data given in Packomania \cite{pack} and include them in our first-principle thermodynamics derivation.

 {
Our model predicts a reduction in NP size with applied magnetic field for materials characterized by negative and positive susceptibilities, which is 
consistent with experimental observations in diamagnetic AgNP \cite{kthiri2021novel} and paramagnetic Nickel NP catalysts  used to grow carbon nanofibers \cite{luo2015strong}. Our theory does not include possible ferromagnetic domains, but we should note that a reduction in size was also reported for ferromagnetic iron NPs \cite{ualkhanova2019influence}.}

Our theory demonstrates the changes in the size and free-energy landscape of AgNPs and shows that, besides the noticeable size effects, a magnetic field acts as an excellent catalyst for nucleation, since it reduces the energy required to grow a stable NP from a metastable state. 

Metastability refers to a non-equilibrium state in which the system resides temporarily before transitioning into a more stable state. A metastable state is a state in which the free energy of the system is at a local minimum of the free energy and not the global minimum at which equilibrium occurs. A system in a metastable state will later evolve to an equilibrium state that corresponds to the global minimum of the free energy \cite{Kalikmanov2013}.

\begin{figure}[ht]
\centering
\includegraphics[width=.8\linewidth]{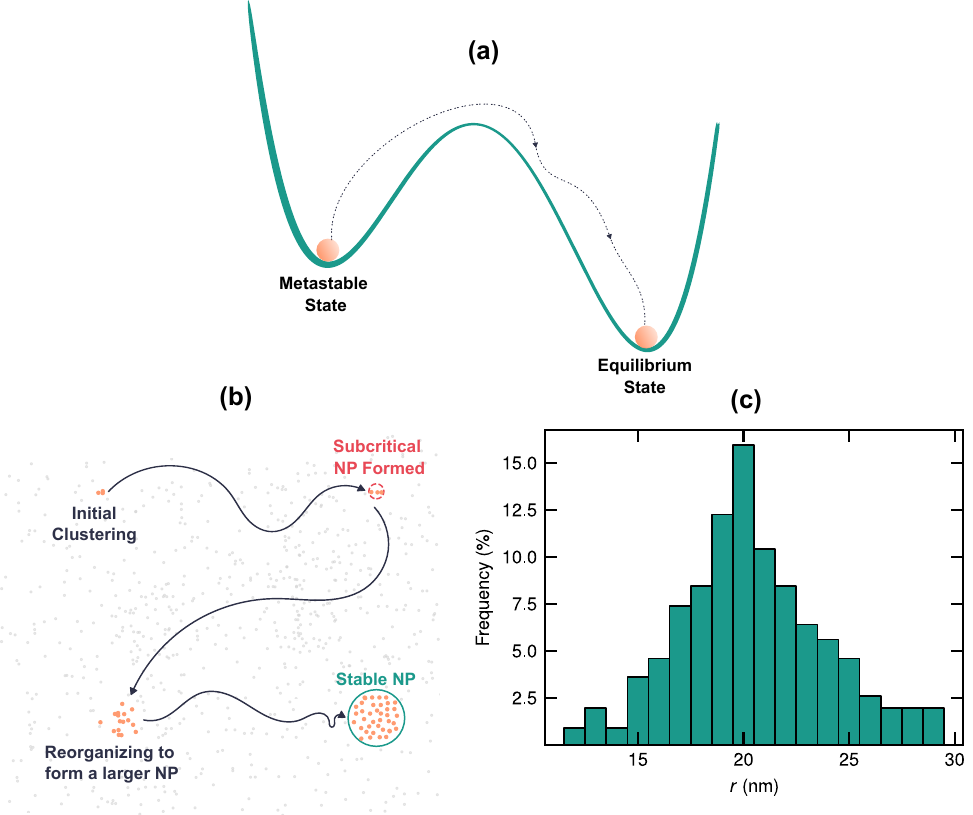}
\caption{
 {(a) Sketch of the Free Energy landscape (b) The formation of NPs (c) Typical NPs radii distribution based on the experiments done by Kthiri et al \cite{kthiri2021novel}.}
}
\label{fig:1}
\end{figure}

 {Figure \ref{fig:1}a Shows the transition from a metastable state to a stable state, the system must overcome an energy barrier, which is the Work of Formation or the minimum amount of work required to transfer a system from the nearest metastable state to the equilibrium state. Here, we consider a distribution of sizes centered on the critical size $n_c$ (Figure \ref{fig:1}c) and not a fixed size due to thermal fluctuations in the system. Two cases apply to sizes outside the Critical Region. For $n < n_c$, where $n$ is the number of atoms in the NP and $n_c$ is its critical size, the particles are subject to Ostwald Ripening \cite{sugimoto2019monodispersed}, which is the dissociation of small NPs in favor of creating larger ones. This occurs because smaller NPs have a large surface area to volume ratio, making it difficult to form particles of that size as shown in Figure \ref{fig:1}b. However, the second case, particles of size $n > n_c$ are considered stable particles because they reside on or around the minimum of the free energy. }
\section*{Results and Discussion}
It has been experimentally observed that exposure of the nucleation medium of different NPs to a magnetic field influences their size and shape \cite{kthiri2021novel, ualkhanova2019influence, luo2015strong}. In this work, we sought to establish a connection between the magnetic field value and the size of NPs, specifically for Silver NPs (AgNPs). This can be achieved by modeling the nucleation barrier and investigating the behavior of the work of formation at its critical point  {$\tfrac{\partial}{\partial x}\Delta\mathcal{F}=0$}. We propose the following expression for the Work of Formation of a NP in the presence of a magnetic field:
 {
\begin{equation}\label{eq:1}
    \Delta \mathcal{F} = -n(x)\bigg(\Delta \mu +\Theta(\sigma)\frac{V_{a}}{2}\frac{3}{\mu_0}\frac{|\chi_m|}{(3+\chi_m)} \mathcal{B}^2\bigg) + 4\pi r^2\gamma
\end{equation}}

Here, $n$ is the number of atoms that make up NP, $\Delta\mu$ is the change in the chemical potential when a stable phase is formed, $\mathcal{B}$ is the external magnetic field measured in Tesla,  $V_a$ is the volume of a single atom $\frac{4}{3}\pi a^3$, $\mu_0$ is the permeability of free space and $\chi_m$ is the magnetic susceptibility. $\gamma$ is the surface free energy and  {$r$ is the radius of the NP formed}. Since the magnetic energy in essence depends on the orientation of the magnetic moment of the NPs with respect to the magnetic field, we consider an additional source of fluctuations in the system that leads to having a distribution of sizes instead of one single size. We infer that these fluctuations arise because the particles do not align completely with the magnetic field. This is represented by the factor  {$\Theta(\sigma) = |\langle\cos(\theta)\rangle|$ is a statistical average over the angle between the external magnetic field and the dipole moment. We weigh the factor $\cos(\theta)$ by a gaussian with a standard deviation $\sigma$, which we express the distribution in terms of. We elaborate further on this average in detail in the Supplementary Information.}

Figure \ref{fig:2} outlines our approach to the geometry of the NPs. We investigate NP formation by considering the 3D packing of small identical spheres of atomic radius $a$ into a larger sphere (NP) of radius  $r$  as shown in Fig. \ref{fig:2}a. We have found analytical approximations for the numerical data for the sphere packing given as open source on packomania\cite{pack}. However, it is more realistic to introduce a physically motivated fit of the following form:

\begin{equation}
    n(x) = \phi_b(x-\delta)^3 + \phi_d[x^3-(x-\delta)^3]
    \label{eq:n}
\end{equation}

Where $\phi_b = \frac{\pi}{3\sqrt{2}}$ is the optimal packing fraction and $\phi_d$ is the defective packing fraction that represents the imperfect packing of atoms on the surface of the NP. The performance of both approaches is shown in Fig. \ref{fig:2}b. The value of $R^2$ is calculated using $R^2 = 1-\tfrac{\sum_i(y_i-\hat{y}_i)^2}{\sum_i(y_i-\bar{y})^2}$ where $y_i$ is the $i$-th data point corresponding to $x_i$ of packomania\cite{pack}, $\bar{y}$ is the mean of the data points from packomania and $\hat{y}_i=f(x_i)$ where $f(x)$ is the fit function. In this treatment, we consider a NP to be an optimally-packed sphere with a thickness surface $\delta$ that is imperfectly packed. We can extract the value of $\phi_d$ for different values of $\delta$ using an ordinary least squares fit employing data from Packomania \cite{pack}. The form of Eq. \eqref{eq:n} is suggested by the Steiner formula\cite{Morvan2008}, which is a result that describes how the volume of a convex body grows when it is uniformly thickened in all directions. The analysis is done with $\delta = 4$ which corresponds to a surface that is two atoms thick. {We note that the particular choice of $\delta$ (and equivalently $\phi_d$) does not have a noticeable effect on the radius-field relation, as shown in the Sensitivity Analysis in the Supplementary Information (S5)}. The Steiner-type fit in Eq. \eqref{eq:n} has a clear advantage over other sphere-packing methods, as shown in \ref{fig:2}c.
 \begin{figure}[hbt!]
    \centering
    \includegraphics[width=.9\linewidth]{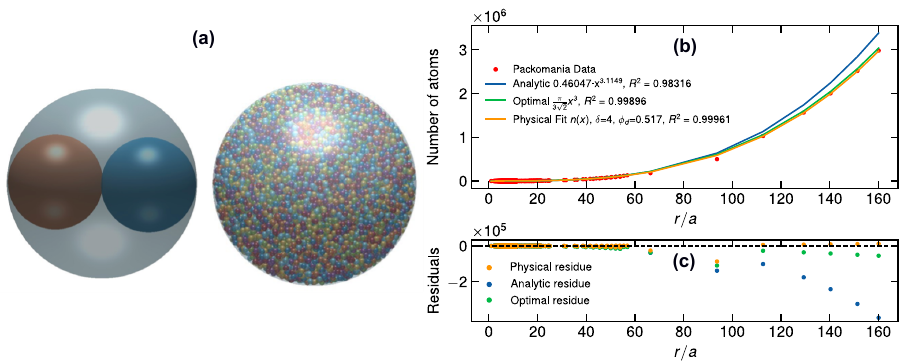}
    \caption{(a) Sphere packing for $r/a=2$ (left) and $r/a = 35$. The $r/a = 2$ diagram is exact, while the diagram $r/a = 35$ is an illustration, obtained by randomly distributing the atoms inside the NP. (b)  {A comparison between the optimal packing fraction, an analytical approximation of packomania's data\cite{pack} and Eq. \eqref{eq:n}. (c) A residuals plot comparing the performance of all fits}.}
    \label{fig:2}
\end{figure}
\newpage
 Finding the critical value of $\Delta\mathcal{F}$  {(where $x = r/a$): 
 \begin{equation}
   \frac{x}{n'(x)}= \bigg[\frac{x_1}{n'(x_1)}+\bigg(\frac{x_1}{n'(x_1)} -\frac{x_2}{n'(x_2)}\bigg)\Theta(\sigma)\bigg(\frac{\mathcal{B}}{\mathcal{B}_2}\bigg)^2\bigg]
   \label{eq:3}
 \end{equation}
}

\begin{figure}[!htbp]
    \centering
    \includegraphics[width=1 \linewidth]{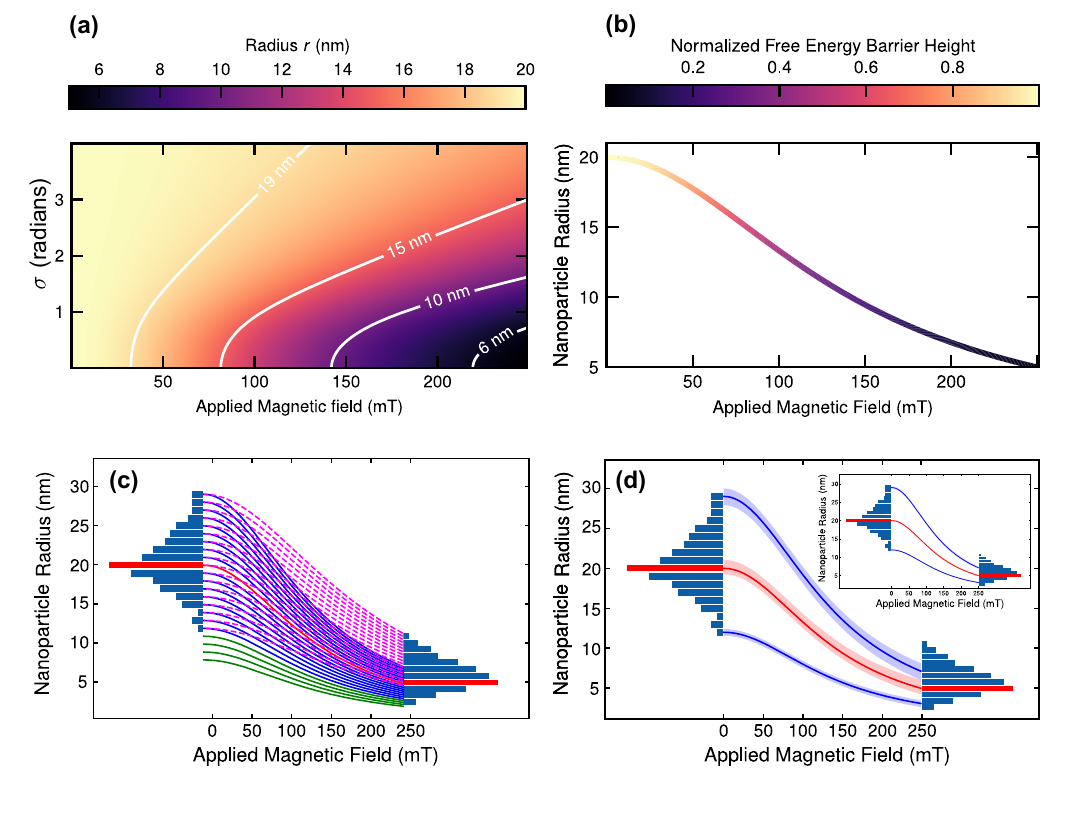}
    \caption{ (a) A heatmap demonstrating the impact of the orientation angle distribution on how the radius transforms under a magnetic field with $\sigma = 0$ rad (deterministic orientation). (b) The field-radius relation from Eq.~\eqref{eq:3} with the respective work of formation barrier height relative to the highest experimental point at $r = 20$ nm, $\mathcal{B}=0$ mT. The normalized Free Energy Barrier Height is $\frac{\Delta\mathcal{F}}{\Delta\mathcal{F}_{max}} $ where $\Delta\mathcal{F}_{max}$ is the Work of Formation from Eq.~\eqref{eq:1} corresponding to the $r=20$ nm, $\mathcal{B}=0$ mT case (c) Comparison between theory and experiments for  AgNPs. The histograms are made with experimental data from Kthiri et al \cite{kthiri2021novel}. 
    The red solid curve contains the two anchor data points from which parameters are extracted, namely $(r_1, \mathcal{B}_1=0$ mT) and $(r_2, \mathcal{B}_2=250$ mT).  The solid curves are calculated with $\sigma = 0$ rad (deterministic orientation) and the dashed with $\sigma=1.2$ rad (Gaussian-averaged orientation). The green curves are for starting points beyond the $\mathcal{B} = 0$ mT histogram {(d) The uncertainty bands for the central and extreme experimental points for $E_s$ (main) and $\Delta\mu$ (inset). The bands are estimated based on a $\pm20\%$ deviation from the fit values obtained via Eqs. \eqref{Es} and \eqref{delta_mu}}.}
    \label{fig:3}
\end{figure}

The prime indicates a derivative with respect to $x$. The pair $(x_1, \mathcal{B}_1=0)$ corresponds to the radius of the NP in the absence of a magnetic field, while $(x_2, \mathcal{B}_2\neq0)$ is the radius of the NP formed at its respective magnetic field $\mathcal{B}_2$.
The solid red curve in Fig.~\ref{fig:3} is obtained with these two values in Eq.~\eqref{eq:3}.  The other curves are calculated with the same equation, taking the center of each histogram at  {$(\bar{x}_1, \mathcal{B}_1=0)$ and calculating the corresponding $(\bar{x}_2, \mathcal{B}_2\neq0)$ through the relation ${x_2/n'(x_2)}=\frac{\bar{x}_2/n'(\bar{x}_2)}{\bar{x}_1/n'(\bar{x}_1)}[{x_1/n'(x_1)}]$}. This shows that our approach is predictive and only needs two points to generate the whole family of curves.

 Figure \ref{fig:3} shows our application of the theory for AgNPs parameters given by Khiri et al \cite{kthiri2021novel}, where $r_1 = 20 \text{ nm}$, $r_2 = 5 \text{ nm}$ and $\mathcal{B}_2 = 250 \text{ mT}$. Figure \ref{fig:3}a is a plot of the field-radius relation Eq. \eqref{eq:3}, we notice a drastic change in Work of Formation from $r = 20$ to $r = 5$ nm.  Furthermore, we can see that the magnetic field reduces the height of the free energy barrier and acts as a catalyst for nucleation by requiring less energy to form a stable NP, provided that nanoscale fluctuations do not dominate the free energy landscape. This indicates that nucleation is more likely to happen in the presence of a magnetic field, which is a bonus over reduction in size. In Figures \ref{fig:3}b and \ref{fig:3}c, we include the effect of the orientation of the magnetic moment of the NP against the magnetic field on the radius of the NP. In Figure \ref{fig:3}c, we generalize the relation to all NP radii obtained by Kthiri et al \cite{kthiri2021novel}.
\newpage

{
The lowest solid blue curve in Fig.~\ref{fig:3}c shows that our results are robust for radii above $\approx3$ nm. From Eq.~\eqref{eq:n}, this corresponds to a NP with ($\approx5740$ atoms). Below this scale non-extensive effects likely dominate \cite{Guisbiers01012019} and the predictions are illustrative only. More details are given in Supplementary Information. Furthermore, Fig.~\ref{fig:3}d shows that fluctuations in $E_s$ (main) have a larger impact than those on $\Delta\mu$ (inset).
}
\newpage 

\subsection*{Limitations and Range of Validity}
 Our model adopts an extensive formulation that assumes additive contributions from the volume and surface terms. Although this treatment is consistent with conventional bulk thermodynamics, finite-size correlations, surface reconstructions, and long-range nanoscale effects may lead to deviations from extensivity. Recent studies \cite{Guisbiers01012019,MANIOTIS2025116285} have highlighted the importance of non-extensive frameworks, such as Tsallis statistics and Hill’s subdivision potential, which incorporate additional terms to capture these correlations. Although such effects lie beyond the present scope, they may become significant for very small NPs, particularly through their influence on the height of the free energy barrier. For example, temperature fluctuations in gold NPs, which are structurally similar to silver, reach $\sim$ 1$\%$ for 3.4 nm particles but increase to $\sim$10$\%$ for 0.7 nm \cite{Guisbiers01012019}.
In our model, these non-extensive effects are reflected qualitatively by the solid green lines in Fig. \ref{fig:3}c, which were manually added since the predictive accuracy decreases in this small size regime. Ferromagnetic domain physics is likewise not included. Quantitative predictions are therefore most reliable for systems where (i) the magnetic response is isotropic and single domain, (ii) particle shapes are near spherical, and (iii) radii remain above the regime where strong non extensive temperature fluctuations are reported. We demonstrate the robustness of the theory against fluctuations in the fit parameters ($\Delta\mu$, $E_s$, $\delta$, $\phi_d$, $\sigma$) in Supplementary Information S5, which confirms its stability against reasonable variations. Barrier heights and extrapolated families in Fig. 3 should thus be viewed as model based estimates rather than direct observables.
A natural bridge to atomistic descriptions lies in the use of cDFT, which treats NPs as continuous density fields rather than sharp hard spheres \cite{Kalikmanov2013}. Although such formulations can relax the capillarity approximation, their mean-field nature limits accuracy in the nanometer regime, where atomic discreteness and quantum correlations dominate. Moreover, the results depend sensitively on the excess free energy functional chosen \cite{PhysRevE.98.012604}. At the opposite end, electronic DFT can capture these correlations but scales poorly with system size \cite{D1RA04876G}: even our smallest NP ($r = 2$ nm) contains $\sim10^3$ atoms, beyond the practical range of DFT ($\sim 100$ atoms). This gives the advantage to our closed-form analytical approach compared to otherwise computationally prohibitive models. 
{
For a recent review highlighting ongoing efforts to bridge these scales through hybrid and multiscale frameworks, see Giovannini et al~\cite{ Giovannini2023-fs}.}
 
\section*{Materials and Methods}
\subsection*{Theoretical Model}
 {
We begin by rewriting Eq. \eqref{eq:1} (which is derived in S1 in the Supplementary Information) as:
\begin{equation}
    \Delta \mathcal{F} = -n(x)\bigg[\Delta\mu+\Theta(\sigma)E_\mathcal{B}\bigg(\frac{\mathcal{B}}{\mathcal{B}_2}\bigg)^2\bigg] + E_sx^2
    \end{equation}
With the following definitions: 
\begin{subequations}
    \begin{align}
        x &\equiv r/a\\
        E_s &\equiv4\pi\gamma  a^2  \\
        E_\mathcal{B} &\equiv \frac{4/3\pi a^3}{2} \frac{3}{\mu_0}\frac{|\chi_m|}{(3+\chi_m)} (\mathcal{B}_2)^2\label{b}
    \end{align}
\end{subequations}
where $\mathcal{B} = \mathcal{B}_1=0 \leftrightarrow r = r_1$ and $\mathcal{B} = \mathcal{B}_2 \leftrightarrow r= r_2$. Now, we maximize $\Delta{\mathcal{F}}$ (at $\sigma=0$) with respect to $x$ to find the dependence of the particle radius on the magnetic field:
\begin{equation}
    \frac{d}{dx} \Delta\mathcal{F} = -n'(x)\bigg[\Delta\mu + E_\mathcal{B}\bigg(\frac{\mathcal{B}}{\mathcal{B}_2}\bigg)^2\bigg]+2E_sx = 0
    \label{dDF}
\end{equation}
Using the experimental points $\mathcal{B} = \mathcal{B}_1=0 \leftrightarrow r = r_1$ and $\mathcal{B} = \mathcal{B}_2 \leftrightarrow r= r_2$, we find:
\begin{subequations}
    \begin{align}
        2E_s = \frac{E_\mathcal{B}}{\big(\tfrac{x_1}{n'(x_1)} -\frac{x_2}{n'(x_2)}\big)}\label{Es}\\
    \Delta\mu=\frac{x_1}{n'(x_1)} \frac{E_\mathcal{B}}{\big(\tfrac{x_1}{n'(x_1)} -\frac{x_2}{n'(x_2)}\big)}\label{delta_mu}
    \end{align}
\end{subequations}}
 {
Now, we substitute the newly found values of $\Delta\mu$ and $E_s$ into equation \eqref{dDF} to find (at any $\sigma$):
\begin{equation}
    \frac{x}{n'(x)}= \bigg[\frac{x_1}{n'(x_1)}+\bigg(\frac{x_1}{n'(x_1)} -\frac{x_2}{n'(x_2)}\bigg)\Theta(\sigma)\bigg(\frac{\mathcal{B}}{\mathcal{B}_2}\bigg)^2\bigg]\equiv Q(\mathcal{B})
\end{equation}
$Q(\mathcal{B})$ is defined as the right-hand side of the previous equation. Solving for $x$, we get:
 \begin{equation}
     x(\mathcal{B})=\frac{\,1-Qa_1 \pm \sqrt{(Qa_1-1)^2-4Q^2 a_2 a_0}\,}{\,2Qa_2\,},
\qquad Q\equiv Q(\mathcal{B})\
\label{rel}
 \end{equation}
 with
\begin{equation}
    a_2=3\phi_b,\qquad a_1=6(-\phi_b+\phi_d)\delta,\qquad a_0=3(\phi_b-\phi_d)\delta^2.
\end{equation}
We take the positive branch of Eq. \eqref{rel} since it is the branch that produces $x(\mathcal{B}=0) = x_1$ and $x(\mathcal{B}=\mathcal{B}_2) = x_2$. We can use the positive branch of Eq. \eqref{rel} to determine the radius of the NP as a function of the applied magnetic field. This equation can be used, provided that the radius $r_2$ corresponding to a certain value of the magnetic field, namely $\mathcal{B}_2$ is known. Although this is useful to model the experiment of Kthiri et al.\cite{kthiri2021novel}, we still need to obtain a relationship between the final radius and the initial radius in order to be able to reproduce the relation for all different radii corresponding to $\mathcal{B} = 0$ as shown in Fig. \ref{fig:3}c.  We start by taking $\tfrac{x_1}{n'(x_1)}$ out of the denominator in Eq. \eqref{delta_mu} to get:}
 {
\begin{equation}
    \frac{\Delta\mu}{E_\mathcal{B}}=\frac{1}{1-\frac{x_2n'(x1)}{x_1n'(x_2)}}
\end{equation}
Solving for $\tfrac{x_2n'(x1)}{x_1n'(x_2)}$:
\begin{subequations}
    \begin{align}
            \frac{E_\mathcal{B}}{\Delta\mu} = 1 - \frac{x_2n'(x_1)}{x_1n'(x_2)} \\
            \frac{x_2n'(x1)}{x_1n'(x_2)}= 1-\frac{E_\mathcal{B}}{\Delta\mu}\equiv \frac{1}{\lambda}\\
    \end{align} 
\end{subequations}
We observe that the RHS of the above equation is a constant. We conveniently choose this constant to be:
\begin{equation}
    \frac{1}{\lambda} = \frac{\bar{x}_1/n'(\bar{x}_1)}{\bar{x}_2/n'(\bar{x}_2)}
\end{equation}
Where the barred points correspond to the highest frequency data points in the work of Kthiri et al\cite{kthiri2021novel}. This gives the final expression:
\begin{equation}    {x_2/n'(x_2)}=\lambda[{x_1/n'(x_1)}] =\frac{\bar{x}_2/n'(\bar{x}_2)}{\bar{x}_1/n'(\bar{x}_1)}[{x_1/n'(x_1)}]
\end{equation}
After solving for $x_2$:
\begin{equation}
    x_2 =
\frac{n'(x_1)}{6\,\phi_b\,\lambda\,x_1}
\Bigg[1
- \frac{6\,\delta\,(\phi_d - \phi_b)\,\lambda\,x_1}{n'(x_1)}
\pm
\sqrt{
\left(
1
- \frac{6\,\delta\,(\phi_d - \phi_b)\,\lambda\,x_1}{n'(x_1)}
\right)^2
+ \frac{36\,\delta^2\,\phi_b\,(\phi_d - \phi_b)\,\lambda^2\,x_1^2}{\big(n'(x_1)\big)^2}
}
\Bigg]
\label{x2}
\end{equation}
Similarly to Eq. \eqref{rel}, we take the positive branch of Eq. \eqref{x2}, since it reproduces the physical result of $x_2 = \bar{x}_2$ at $x_1 = \bar{x}_1$.
We can also look at the asymptotic behavior of both branches as $x_1\to\infty$, by multiplying $n'(x_1)/x_1$ into the bracket, we see that the positive branch $x^{(+)}_2 \sim (x_1-\text{constant})$ and $x^{(-)}_2\sim \frac{1}{x_1}$.}  {More details of the derivation of these equations are given in Supplementary Information (S3-S4). }
The number of Packomania points used to fit $\phi_b$ and $\phi_d$ is 1,045; $R^2$ and residuals are reported in Fig.~\ref{fig:2}.

{A sensitivity analysis summarizing all parameters used in the fits, together with their physical units, best-fit values, CI 95$\%$, and how each was determined (either from experimental anchor points or the Packomania dataset) is given in Section S5 of the Supplementary Information.}

\section*{Conclusion}
In summary, we developed a sphere-packing based classical nucleation model that quantitatively links NPs size to the applied magnetic field. Calibrated to two experimental references in AgNPs, the model reproduces the observed reduction from ~20 nm at $\mathcal{B}=0$ to $\sim5$ nm at $\mathcal{B}=250$ mT and reveals a field induced reduction and broadening of the nucleation barrier. The analysis shows that magnetic fields can act as effective catalysts for nucleation, shifting the most-probable size toward smaller radii and broadening size distributions under suitable thermal conditions. The robustness of the predictions to parameter variations and angular disorder is detailed in Section S5 (SI), and possible non-extensive effects at the smallest sizes are discussed in the main text. Overall, these results provide a consistent quantitative framework linking NPs size to applied magnetic field.

\section*{Author contributions statement}
M.F.P. coordinated the driving theory project, derived Eq. \eqref{eq:3} and generalized it to all radii as in Fig. \ref{fig:3}, and secured funding. Y.T. derived the theory from first-principles thermodynamics, derived the magnetic free energy expression and produced all figures. M.G. extracted data from experiments in the literature and wrote the code that led to Fig. \ref{fig:2}a. All authors contributed to writing and reviewing the manuscript.

\section*{Funding}
This publication is based upon work supported by Khalifa University under Award No. CIRA-2021-108.
\section*{Competing interests}
The authors declare no competing interests.
\section*{Data availability}
The datasets used and/or analyzed during the current study are available from the corresponding author on reasonable request.

\section*{Appendix / Supplementary Information}

\renewcommand{\theequation}{S\arabic{equation}}
\setcounter{equation}{0}
\renewcommand{\thefigure}{S\arabic{figure}}
\setcounter{figure}{0}

\renewcommand{\thetable}{S\arabic{table}}
\setcounter{table}{0}
\noindent 
\section*{S1. Work of Formation}

In order to find an analytic form for the Work of Formation, we consider a vapor of $N$  atoms under constant volume and temperature. The Helmholtz Free Energy reads:

 \begin{equation}\label{eq:S1}
      \mathcal{F} = N \mu - PV
 \end{equation}

Here, $\mu$ is the chemical potential of the system, $P$ is the pressure, and $V$ is the volume occupied by the vapor. Next, we consider the spontaneous creation of a Nanoparticle (NP) of size $n$ as a result of a few atoms transitioning from a metastable state. Physically speaking, the creation of a NP of size $n$ is equivalent to introducing a surface into the nucleation medium. This surface is considered an energy barrier that prevents the flow of atoms out of it and into the radius shell $r$, surrounding them as shown in Figure 1b of the manuscript.

Next, we show how the expression of the free energy of the system in Eq. \eqref{eq:S1} changes as a NP is spontaneously formed. The cost of forming a new NP of size $n$ is $n \mu$. This will be split into two parts; the first is the chemical potential contribution of the new phase $n\mu_{\text{NP}}$ and the second is the cost of creating the surface. Intuitively, the energy of the NP surface must depend on its size, therefore, we define the surface energy as $E_s = \gamma A_s$. Here, $\gamma$ is a proportionality factor and $A_s$ is the surface area of a NP of size $n$. The final expression of $\mathcal{F}$ after forming a single NP is: \begin{equation}\label{eq:2}
    \mathcal{F}_{\text{final}}=(N-n)\mu -  PV+n\mu_{\text{NP}} + \gamma A_s
\end{equation}

Finally, we consider the change in the free energy which corresponds to the Work of Formation of a NP of size $n$
\begin{equation}
    \begin{aligned}
    \Delta \mathcal{F} &= \mathcal{F}_{\text{final}}-\mathcal{F}\\&= n(\mu_{\text{NP}}-\mu) + \gamma A_s\\
     &= -n\Delta\mu + \gamma A_s, &&
\end{aligned}
\end{equation}
where $\Delta\mu=\mu-\mu_{\text{NP}}$. This form of free energy matches the capillarity approximation \cite{gibbs}, which is a special case of a more general expression derived by McClurg and Flagan \cite{mcclurg}. Physically, $\gamma$ is the surface free energy, which can be obtained using different computational and experimental methods \cite{aqra2014surface, fox,zenkiewicz2007methods, kwok1999contact}. The surface area $A_s$ can take multiple forms depending on the geometry of the NPs. The relevant value for this work is $A_s=4\pi r^2$ since we are considering spherical geometry.   {Note that this formulation assumes non-extensive effects on the nanoscale, which is addressed in the main text with more details.}

\section*{S2. Magnetic Free Energy}
First we define the magnetic field and the relationship between the total magnetization as:

\begin{subequations}
    \begin{equation}\label{eq:4a}
    \vec{\mathcal{B}} = \mu_0(\vec{\mathcal{H}} + \vec{\mathcal{M}})
    \end{equation}
    \begin{equation}\label{eq:4b}
          {
        \vec{\mathcal{M}} = \frac{3}{\mu_0}\frac{\chi_m}{3+\chi_m}\vec{\mathcal{B}}}
    \end{equation}
    \begin{equation}\label{eq:4c}
        \vec{\mathcal{B}} = \mu_0(1+\chi_m)\vec{\mathcal{H}}
    \end{equation}
\end{subequations}\\
Where $\chi_m$ is   {the bulk constant susceptibility of the material that is} assumed to be constant since the NPs are said to be spherical.   {Additionally, the expression for magnetization is adjusted for the geometry of the NP, which we consider a magnetized sphere in an external field as per Griffiths and Jackson \cite{grif,jackson1999classical}.} 

For a magnetic field $\vec{\mathcal{B}} = \mathcal{B}\hat{z}$, we shall assume that the thermal fluctuations in the system cause $\vec{\mathcal{M}}$ to become disoriented from the $\hat{z}$ direction, hence inducing an external magnetization $\delta\vec{\mathcal{M}_\perp}$ that leads to $\vec{\mathcal{M}} \equiv |\mathcal{M}|\cos\theta \hat{z}+\delta\vec{\mathcal{M}}_{\perp}$, and consequently $\vec{\mathcal{H}} \equiv |\mathcal{H}|\cos\theta \hat{z}+\delta\vec{\mathcal{H}}_{\perp}$.\\
\\
The perpendicular components induce their own magnetic field $\delta\vec{\mathcal{B}}_{\perp}$ such that 
\begin{equation}
    \delta\vec{\mathcal{B}}_{\perp} = \mu_0(\delta\vec{\mathcal{H}}_{\perp}+\delta\vec{\mathcal{M}}_{\perp})
\end{equation}
This allows us to rewrite Eq. \eqref{eq:4a} as:
\begin{equation}\label{delB}
    \vec{\mathcal{B}} + \delta\vec{\mathcal{B}_{\perp}} = \mu_0(\mathcal{H}\cos\theta \hat{z}+\delta\vec{\mathcal{H}}_{\perp} + \mathcal{M}\cos\theta \hat{z}+\delta\vec{\mathcal{M}}_{\perp})  
\end{equation}
Equation \eqref{delB} is the general relation that shows the balance between the magnetization induced by the external field and the corresponding loss due to thermal fluctuations in the system. In order to extract an expression for the magnetic free energy, we start with the expression for the internal energy in a magnetic system by Callen \cite{callen_thermodynamics_1985}. The empty solenoid energy, which is not thermodynamically significant, is conveniently absorbed into $\mathcal{U}$.

\begin{equation}\label{eq:7}
    d\mathcal{U} = TdS - PdV + \mu dN + \vec{\mathcal{B}}\cdot d\vec{m}
\end{equation}
Here, $\vec{m}\equiv V_{NP}\vec{\mathcal{M}}$ is the magnetic moment of the system and $V$ is the volume of the NP. We rewrite Eq. \eqref{eq:7} as:
\begin{equation}\label{eq:8}
    d\mathcal{U} = TdS - PdV + \mu dN + \vec{\mathcal{B}}\cdot d(V_{NP}\vec{\mathcal{M}})
\end{equation}
  {To evaluate the magnetic free energy, we perform a Legendre transform over the thermal and magnetic degrees of freedom. In particular, the externally applied magnetic field $\vec{\mathcal{B}}$, is chosen as the control variable, since the magnetic moment is subject to noise-induced fluctuations.} Therefore, we define $\mathcal{F} \equiv \mathcal{U}-TS - \vec{\mathcal{B}}\cdot (V_{NP}\vec{\mathcal{M}})$ to obtain:
\begin{equation}
    d\mathcal{F} = -SdT-PdV-V_{NP}\vec{\mathcal{M}}\cdot d\vec{\mathcal{B}} + \mu dN
\end{equation}
Using Eq. \eqref{eq:4b}, we get:
\begin{equation}
    d\mathcal{F} = -SdT-PdV-V_{NP}\frac{3}{\mu_0}\frac{\chi_m}{3+\chi_m}\cos\theta\mathcal{B} d\mathcal{B} + \mu dN
\end{equation}
At constant $T,V,N$: 
  {
\begin{equation}
    \bigg(\frac{d\mathcal{F}}{d\mathcal{B}}\bigg)_{T,V,N} = -V_{NP}\frac{3}{\mu_0}\frac{\chi_m}{(3+\chi_m)}\cos\theta \mathcal{B}
\end{equation}}
Integrating over the measure $d\mathcal{B}$, we obtain the magnetic free energy:
  {
\begin{equation}\label{FB}
    \mathcal{F}_{mag}=-\frac{V_{NP}}{2}\ \frac{3}{\mu_0}\frac{\chi_m}{(3+\chi_m)}\cos\theta \mathcal{B}^2
\end{equation}}
  {Our formalism uses the magnetic field $\mathcal{\vec{B}}$, which is measured in Tesla (\text{T})\cite{247071}, for a direct comparison with the experiments by Kthiri et al\cite{kthiri2021novel}. An alternative treatment using the magnetic field strength $\mathcal{\vec{H}}$ measured in Ampere per meter\cite{247071} $\text{A/m}$ is possible by direct substitution of Eq. \eqref{eq:4c} into Eq. \eqref{FB}.
}

  {Due to the orientation factor in this formalism, the negative sign of the magnetic free energy is preserved whether we are discussing a diamagnetic system or a paramagnetic system, This allows us to rewrite the magnetic free energy as:}
\begin{equation}
    \mathcal{F}_{mag}=-\Theta(\sigma)\frac{V_{NP}}{2}\ \frac{3}{\mu_0}\frac{|\chi_m|}{(3+\chi_m)} \mathcal{B}^2
\end{equation}
  We use $V_{NP}=nV_a$ in the main text, with $V_a$ being the atomic volume and $n$ being the number of atoms that make up the NP. To capture the statistical nature of the system, we introduce $\Theta(\sigma)\equiv |\langle\cos\theta\rangle|$. The average is taken over $[0,2\pi]$ and weighed by a Gaussian centered at $\theta=\pi$, since silver is diamagnetic, with standard deviation $\sigma$:  
\begin{equation}
    |\langle \cos\theta\rangle|
=\left\lvert\frac{\displaystyle\int_{0}^{2\pi}\cos\theta\,
\exp\!\Big[-\tfrac{(\theta-\pi)^2}{2\sigma^2}\Big]\,d\theta}
{\displaystyle\int_{0}^{2\pi}
\exp\!\Big[-\tfrac{(\theta-\pi)^2}{2\sigma^2}\Big]\,d\theta}\right\rvert
\qquad (0\le \theta\le 2\pi).
\end{equation}

\begin{figure}
    \centering
    \includegraphics[width=1\linewidth]{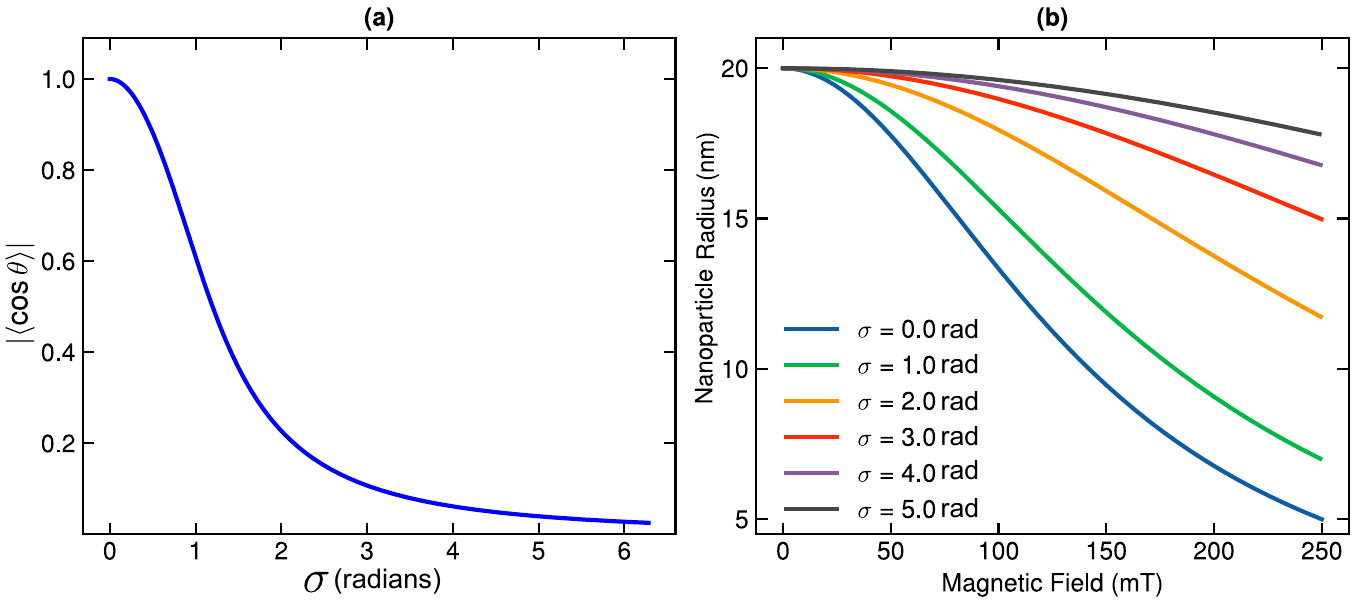}
    \caption{(a) The change in the average of the angle factor with respect to $\sigma$ (b) The effect of the orientation factor on the field-radius relationship.}
    \label{fig:s1}
\end{figure}

This expression does not affect the mechanical work degree of freedom and works in the case of an isobaric ($N,P,T$) system, with the Gibbs Free Energy as its thermal potential. For instance (through a similar process as the previous): 
\begin{equation}
    \begin{aligned}
        \mathcal{G} &= \mathcal{U}-TS - \vec{\mathcal{B}}\cdot (V_{NP}\vec{\mathcal{M}})+PV\\
        d\mathcal{G}&= -SdT +VdP-V_{NP}\vec{\mathcal{M}}\cdot d\vec{\mathcal{B}} + \mu dN\\
        d\mathcal{G}&= -SdT +VdP-V_{NP}\frac{3}{\mu_0}\frac{\chi_m}{3+\chi_m}\cos\theta\mathcal{B} d\mathcal{B} + \mu dN\\
        \mathcal{G}_{mag}&=-\Theta(\sigma)\frac{V_{NP}}{2}\ \frac{3}{\mu_0}\frac{|\chi_m|}{(3+\chi_m)} \mathcal{B}^2 =\mathcal{F}_{mag}      
    \end{aligned}
\end{equation}

  {The negative sign of the magnetic term in the Work of Formation is preserved for both diamagnetic and paramagnetic materials. This is consistent with the experimentally observed reduction in dimensions under an applied magnetic field as reported in diamagnetic silver by Kthiri et al \cite{kthiri2021novel}, and for paramagnetic Nickel NP catalysts \cite{luo2015strong}.  
The presence of magnetic domains makes the analysis of ferromagnetic materials beyond the scope of our approach. However, a reduction in size has also been reported in ferromagnetic Iron NPs \cite{ualkhanova2019influence}.  For a diamagnetic material such as silver, the susceptibility $\chi_m=-2.31\times10^{-5}$    is weak, which leads to $\mathcal{F}_{mag}\approx -\Theta(\sigma)\frac{V_{NP}}{2\mu_0}|\chi_m|\mathcal{B}^2$
. In contrast, for a paramagnetic material with a strong isotropic susceptibility, $\frac{|\chi_m|}{3+\chi_m}\approx1$ and $\mathcal{F}_{mag}\approx -\Theta(\sigma)\frac{3V_{NP}}{2\mu_0}\mathcal{B}^2$. Thus, while the qualitatively behavior is  similar for diamagnetic and paramagnetic materials, the larger magnetic free energy for systems of high susceptibility, makes the magnetic field more effective in our model.
}

\section*{S3. Sphere Packing}
  {Literature reports have observed that larger NPs often accumulate more surface defects during growth\cite{https://doi.org/10.1002/admi.202100867}. This tendency motivates the incorporation of a size-dependent defect density in our modeling. To quantify this behavior, we separate the contributions into:
\begin{enumerate}
    \item \textbf{Volume packing}: atoms arranged in the bulk interior.
    \item \textbf{Surface packing}: atoms accommodated at the outermost shell, where geometric packing is less efficient.
\end{enumerate}
For the interior, we assumed ideal close-packing of spheres. The effective bulk packing density is taken as
\[
\phi_b = \frac{\pi}{3\sqrt{2}} \approx 0.7405,
\]
which corresponds to the maximum achievable packing fraction for identical spheres.  
The number of atoms in the bulk is therefore estimated as:
\[
n_{\mathrm{bulk}}(r) = \phi_b \, \frac{V(r - \delta)}{V_a},
\]
where $V(r-\delta)$ is the volume of a reduced sphere of radius $r-\delta$, $V_a$ is the atomic volume, and $\delta$ is the thickness of the surface. At the boundary, atoms cannot be packed with the same efficiency as a result of curvature and incomplete coordination. We therefore introduce a defective surface packing fraction $\phi_d$ that we fit using $1,045$ data points provided by packomania\cite{pack}. This accounts for the imperfect packing at the NP surface. The analysis is done with $\delta = 4$ which corresponds to a surface that is two atomic layers thick. The effect of varying the thickness on the packing fraction is shown in Fig. \ref{fig:s2}b. The effect on the radius is discussed in S5: Sensitivity Analysis.
The number of surface atoms is given by
\[
n_{\mathrm{surf}}(r) = \phi_d \, \frac{V(r) - V(r - \delta)}{V_a}.
\]
And finally, the total atom count is (where $x = r/a$):
\begin{equation}
    n(x) = \phi_b \,(x-\delta)^3 + \phi_d \,\big(x^3 - (x-\delta)^3\big)
\end{equation}
\begin{figure}
    \centering
    \includegraphics[width=1\linewidth]{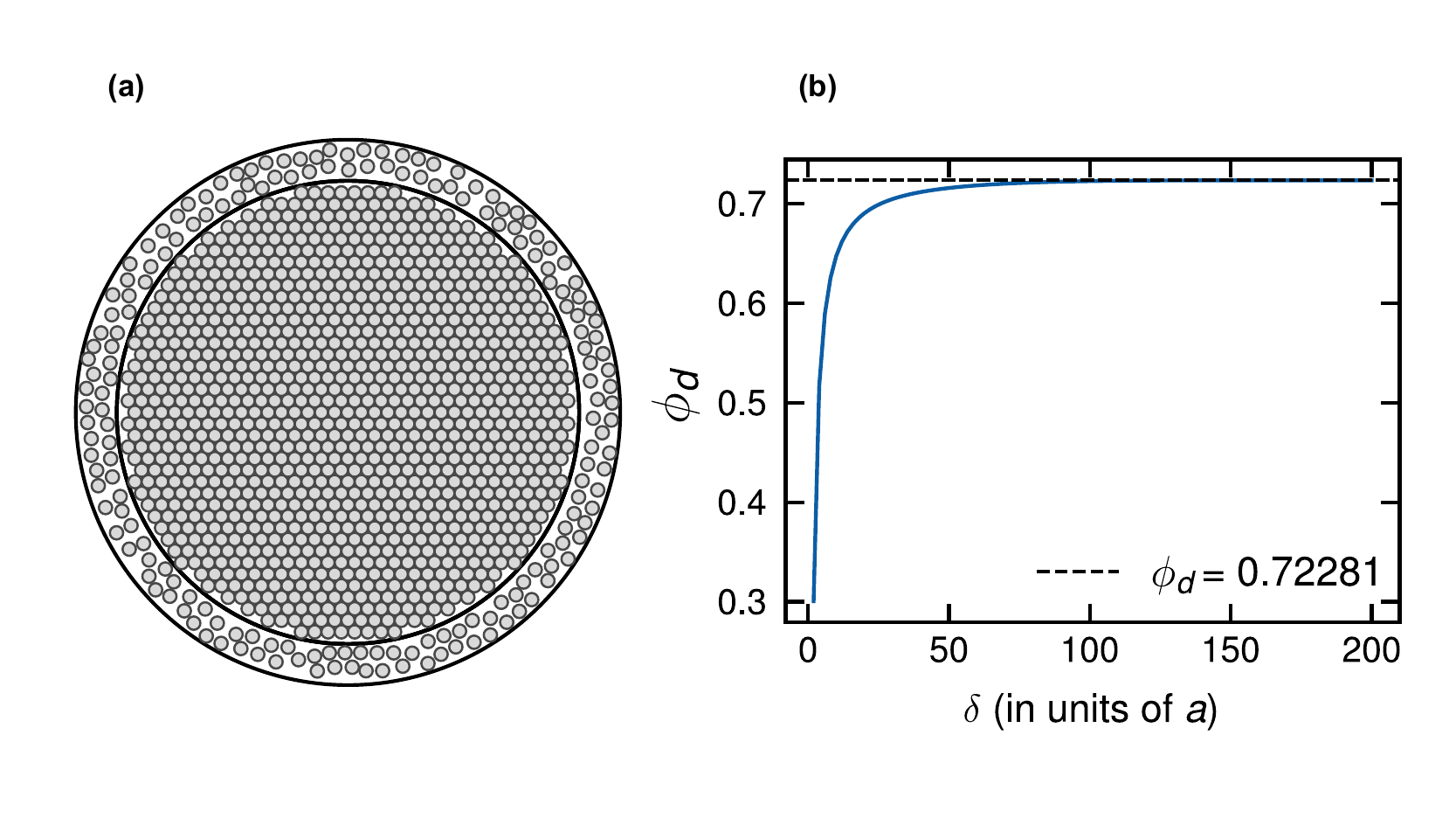}
    \caption{(a) A demonstration diagram of a NP with a perfectly packed core and an imperfectly packed surface of thickness $\delta$ (b) The effect of the surface thickness $\delta$ on the packing fraction $\phi_d$}
    \label{fig:s2}
\end{figure}
We use the formula above, along with $1,045$ data points publicly available from Packomania\cite{pack} to fit the value of $\phi_d$ for any $\delta$.
This choice is an iteration Steiner formula for convex bodies \cite{Morvan2008}, which expresses how the volume of a convex body increases when you thicken it uniformly in all directions.}
\section*{S4. Radius-Magnetic Field Relation}
  {
In order to establish a complete relation between the radius of the NP and the magnetic field, we need to solve for the two unknowns in the work of formation expression:
\begin{equation}\label{eq:15}
    \Delta \mathcal{F} = -n(x)\bigg(\Delta \mu + \frac{V_a|\chi_m|}{2\mu_0(1+\chi_m)}\Theta(\sigma)\mathcal{B}^2\bigg) +  4\pi r^2\gamma
\end{equation}
We rewrite the previous expression as:
\begin{equation}
    \Delta \mathcal{F} = -n(x)\bigg[\Delta\mu+\Theta(\sigma)E_\mathcal{B}\bigg(\frac{\mathcal{B}}{\mathcal{B}_2}\bigg)^2\bigg] + E_sx^2
    \end{equation}
With the following definitions: 
\begin{subequations}
    \begin{align}
        x &\equiv r/a\\
        E_s &\equiv4\pi\gamma  a^2  \\
        E_\mathcal{B} &\equiv \frac{4/3\pi a^3}{2} \frac{3}{\mu_0}\frac{|\chi_m|}{(3+\chi_m)} (\mathcal{B}_2)^2
    \end{align}
\end{subequations}
}
where $\mathcal{B}_2$ is the largest applied magnetic applied to the system. 
We maximize $\Delta{\mathcal{F}}$  with respect to $x$ to find the dependence of the particle radius on the magnetic field:

  {\begin{equation}
    \frac{d}{dx} \Delta\mathcal{F} = -n'(x)\bigg[\Delta\mu + E_\mathcal{B}\Theta(\sigma)\bigg(\frac{\mathcal{B}}{\mathcal{B}_2}\bigg)^2\bigg]+2E_sx = 0 \label{S20}
\end{equation}
This expression is valid for all values of $\mathcal{B}$, x, and $\sigma$. For $\sigma=0$, $\mathcal{B} = \mathcal{B}_1=0 \leftrightarrow r = r_1$ and $\mathcal{B} = \mathcal{B}_2 \leftrightarrow r= r_2$
We evaluate Eq. \eqref{S20} at $(r_1, \mathcal{B}_1)$ to obtain:
}
  {
\begin{subequations}
    \begin{align}
        n'(x_1) \Delta\mu &= 2 E_sx_1\\
        \Delta\mu &= \frac{2E_sx_1}{n'(x_1)} \label{mu_1}
    \end{align}
\end{subequations}
and for $(r_2, \mathcal{B}_2)$:
\begin{subequations}
    \begin{align}
        \Delta\mu + E_\mathcal{B}&=\frac{2E_sx_2}{n'(x_2)}\label{21a}\\
        \Delta\mu &=\frac{2E_sx_2}{n'(x_2)}-E_\mathcal{B}\label{mu_2}
    \end{align}
\end{subequations}
Equating Eqs. \eqref{mu_1} and \eqref{mu_2} and solving for $E_s$:
\begin{subequations}
    \begin{align}
      \frac{2E_sx_1}{n'(x_1)} =\frac{2E_sx_2}{n'(x_2)}-E_\mathcal{B} \\2E_s\bigg(\frac{x_2}{n'(x_2)} -\frac{x_1}{n'(x_1)}\bigg)=E_\mathcal{B}\\
      2E_s = \frac{E_\mathcal{B}}{\big(\tfrac{x_2}{n'(x_2)} -\frac{x_1}{n'(x_1)}\big)}\label{S:Es}
\end{align}
\end{subequations}
Now, we use Eqs. \eqref{mu_1} and \eqref{S:Es} to solve for $\Delta\mu$:
\begin{equation}
    \Delta\mu=\frac{x_1}{n'(x_1)}
\frac{E_\mathcal{B}}{\big(\tfrac{x_2}{n'(x_2)} -\frac{x_1}{n'(x_1)}\big)}
\end{equation}
Finally, we substitute the newly found values of $\Delta\mu$ and $E_s$ into the equation \eqref{dDF}:
\begin{equation}
    n'(x)\bigg[\frac{x_1}{n'(x_1)}
\frac{E_\mathcal{B}}{\big(\tfrac{x_2}{n'(x_2)} -\frac{x_1}{n'(x_1)}\big)} + E_\mathcal{B}\Theta(\sigma)\bigg(\frac{\mathcal{B}}{\mathcal{B}_2}\bigg)^2\bigg]=\frac{E_\mathcal{B}}{\big(\tfrac{x_2}{n'(x_2)} -\frac{x_1}{n'(x_1)}\big)}x  
\end{equation}
Simplifying:
\begin{equation}
    \frac{x}{n'(x)}= \bigg[\frac{x_1}{n'(x_1)}+\bigg(\frac{x_2}{n'(x_2)} -\frac{x_1}{n'(x_1)}\bigg)\Theta(\sigma)\bigg(\frac{\mathcal{B}}{\mathcal{B}_2}\bigg)^2\bigg]
\end{equation}
We can rewrite the expression above as:
\begin{equation}
\frac{x}{n'(x)} = Q(\mathcal{B}),
\qquad
n'(x)=3\phi_b(x-\delta)^2+3\phi_d\!\big(x^2-(x-\delta)^2\big),
\end{equation}
define the field–dependent right–hand side
\begin{equation}
Q(\mathcal{B})\equiv \frac{x_1}{n'(x_1)} 
+ \Bigg(\frac{x_2}{n'(x_2)}-\frac{x_1}{n'(x_1)}\Bigg)\Theta(\sigma)\!\left(\frac{\mathcal{B}}{\mathcal{B}_2}\right)^{\!2}.
\end{equation}
The limiting cases $x(0)=x_1$ and $x(B_2)=x_2$ are directly recovered for $\sigma$=0. Expanding $n'(x)$,
\begin{align}
x^2-(x-\delta)^2 &= 2x\delta-\delta^2,\\
n'(x) &= 3\!\left[\phi_b x^2 + 2(\!-\phi_b+\phi_d)\,\delta\,x + (\phi_b-\phi_d)\delta^2\right]
      \equiv a_2 x^2 + a_1 x + a_0,
\end{align}
with
\begin{equation}
a_2=3\phi_b,\qquad a_1=6(-\phi_b+\phi_d)\delta,\qquad a_0=3(\phi_b-\phi_d)\delta^2.
\end{equation}
Substituting into $x=Q(B)\,n'(x)$ yields a quadratic in $x$,
\begin{equation}
\big(Qa_2\big)x^2+\big(Qa_1-1\big)x+\big(Qa_0\big)=0,
\end{equation}
whose two branches are
\begin{equation}
x(\mathcal{B})=\frac{\,1-Qa_1 \pm \sqrt{(Qa_1-1)^2-4Q^2 a_2 a_0}\,}{\,2Qa_2\,},
\qquad Q\equiv Q(\mathcal{B}).
\label{S:rel}
\end{equation}
\begin{figure}[!h]
    \centering
    \includegraphics[width=1\linewidth]{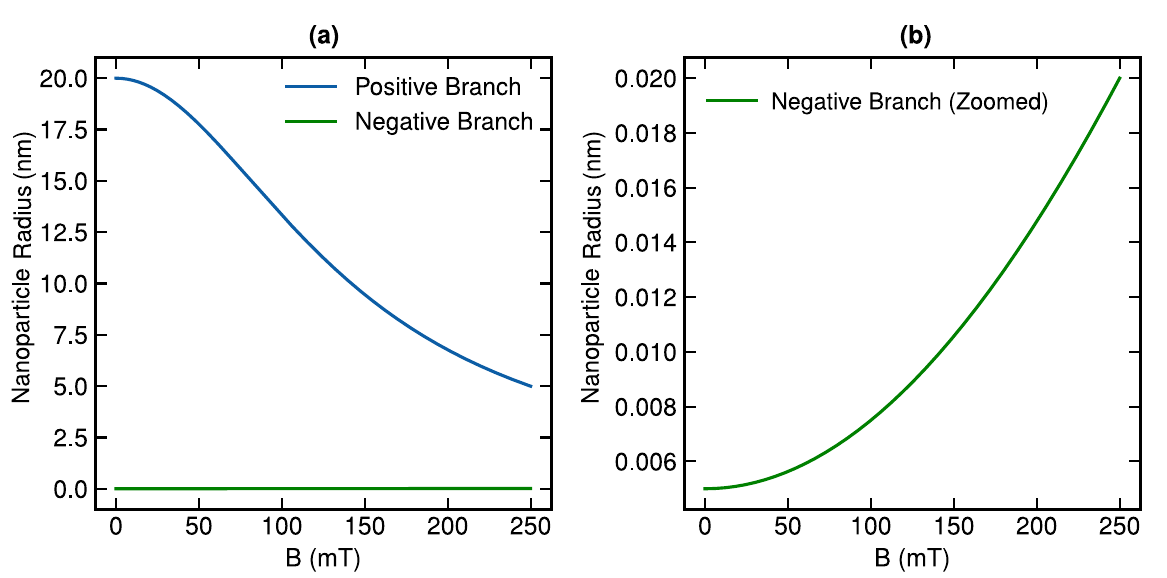}
    \caption{(a) The behavior of the positive and negative branches in Eq. \eqref{S:rel}} (b) Zoom into the unphysical negative branch.
    \label{fig:S3}
\end{figure}
Figure \ref{fig:S3} shows the behavior of both branches of Eq. \eqref{S:rel}. The positive branch recovers the boundary conditions $x(0)=x_1$ and $x(\mathcal{B}_2)=x_2$. The negative branch gives unrealistically small radii that increase as we increase $\mathcal{B}$ and is discarded. We can use Eq.\eqref{S:rel} to determine the radius of the NP as a function of the applied magnetic field. This equation can be used, provided that the radius $r_2$ corresponding to a certain value of the magnetic field, namely $\mathcal{B}_2$ is known. While this is useful to model the experiment of Kthiri et al\cite{kthiri2021novel}, we still need to obtain a relationship between the final radius and the initial radius in order to be able to reproduce the relation for all different radii corresponding to $\mathcal{B} = 0$ as shown in Fig. 3c in the paper.  We start by dividing Eq. \eqref{mu_1} by Eq. \eqref{21a} to get:
\begin{equation}
    \frac{x_1/n'(x_1)}{x_2/n'(x_2)}=\frac{\Delta\mu}{\Delta \mu+E_\mathcal{B}} \equiv\frac{1}{\lambda
}\end{equation}
We observe that the RHS of the equation above is a constant. We conveniently choose this constant to be:
\begin{equation}
    \frac{1}{\lambda
} = \frac{\bar{x}_1/n'(\bar{x}_1)}{\bar{x}_2/n'(\bar{x}_2)}
\end{equation}
Where the barred points correspond to the highest frequency data points in the work of Kthiri et al\cite{kthiri2021novel}, which gives the final expression:
\begin{equation}    {x_2/n'(x_2)}=\frac{\bar{x}_2/n'(\bar{x}_2)}{\bar{x}_1/n'(\bar{x}_1)}[{x_1/n'(x_1)}]
\end{equation}
Which we can solve to find:
\begin{equation}
    x_2 =
\frac{n'(x_1)}{6\,\phi_b\,\lambda\,x_1}
\Bigg[1
- \frac{6\,\delta\,(\phi_d - \phi_b)\,\lambda\,x_1}{n'(x_1)}
\pm
\sqrt{
\left(
1
- \frac{6\,\delta\,(\phi_d - \phi_b)\,\lambda\,x_1}{n'(x_1)}
\right)^2
+ \frac{36\,\delta^2\,\phi_b\,(\phi_d - \phi_b)\,\lambda^2\,x_1^2}{\big(n'(x_1)\big)^2}
}
\Bigg]\label{eq:x2si}
\end{equation}
\begin{figure}[!h]
    \centering
    \includegraphics[width=\linewidth]{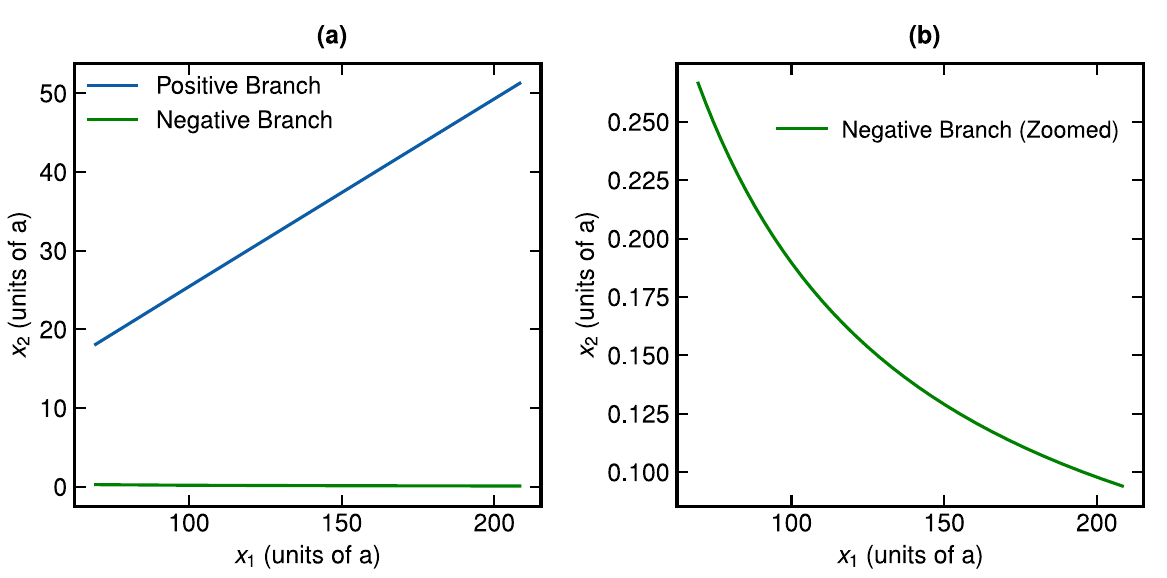}
    \caption{(a) The behavior of the positive and negative branches in Eq. \eqref{eq:x2si} (b) Zoom into the unphysical negative branch.}
    \label{fig:s4}
\end{figure}
Figure \ref{fig:s4} shows the behavior of both branches of Eq. \eqref{eq:x2si} in a similar fashion to Fig. \ref{fig:S3}. We can see that the negative branch values are smaller than $a$, therefore they are nonphysical. Additionally, the negative branch grows smaller as $x_1\to\infty$ which is more reason to discard the branch considering that this behavior is not justifiable experimentally.}
\section*{S5. Sensitivity Analysis}
In this section, we examine the variations in the set of fit parameters $(\Delta\mu, E_s, \delta, \phi_b,\phi_d, \sigma)$ that may affect the conclusions of the theory. We take a local one-at-a-time (OAT) approach where we perturb $\Delta\mu$ and $E_s$ by $\pm20\%$ to see how sensitive the final radius is to changes in them, and finally for $\sigma$ and $\delta$.

  {Variations in the thermodynamic quantities $\Delta\mu$ and $E_s$ (or equivalently $\gamma$) have a modest effect on the outcome compared to the experimental results, with a deviation capped at about $15\%$ for a deviation of $20\%$ from the free energy of the base value of $\gamma$, which is larger than the deviation we observe for the chemical potential. This indicates that surface effects relatively dominate the nucleation landscape. However, the shift of about $15\%$ suggests that the model is robust and a moderate uncertainty does not alter the predictions of the model.}

  {On the other hand, we see a substantial effect due to the variations in $\sigma$. However, this effect is expected since a very large value of $\sigma$ indicates that the NP is not affected by the presence of the magnetic field. It is good to note here that the model considers an average $1:1$ correspondence between the size of a NP in the presence and absence of a magnetic field for a given $\sigma$. This is not necessarily true and can be clearly reflected in the data in Table \ref{table:1}. For instance, it is theoretically possible for a $20$ nm particle to be mapped to any size $r\in [5\text{ nm}, 20\text{ nm}]$ depending on the magnetic moment of the NP.}
\begin{table}[ht]
\centering
\setlength{\tabcolsep}{3pt}        
\renewcommand{\arraystretch}{0.9}   
\footnotesize   
\begin{tabular}{@{}cccccccccccc@{}}
\toprule
$E_s$ (J)    & $r$ (nm) & $\frac{|r-r_2|}{r_2}$ (\%) & $\Delta\mu$ (J) & $r$ (nm) & $\frac{|r-r_2|}{r_2}$ (\%) & $\sigma$ & $r$ (nm)  & $\frac{|r-r_2|}{r_2}$ (\%) & $(\delta, \phi_d)$ & $r$ (nm) & $\frac{|r-r_2|}{r_2}$ (\%) \\ \midrule
4.329638e-27 & 4.248498 & 15.03                      & 2.855453e-29    & 5.234132 & 4.68                       & 0.0      & 5.000000  & 0.00                       & (2.0, 0.3006)      & 5.0      & 0                          \\
4.437879e-27 & 4.327065 & 13.46                      & 2.926839e-29    & 5.209676 & 4.19                       & 0.5      & 5.461417  & 9.23                       & (3.6, 0.4932)      & 5.0      & 0                          \\
4.546120e-27 & 4.404841 & 11.90                      & 2.998225e-29    & 5.185461 & 3.71                       & 1.0      & 6.988944  & 39.78                      & (5.2, 0.5673)      & 5.0      & 0                          \\
4.654361e-27 & 4.481837 & 10.36                      & 3.069612e-29    & 5.161483 & 3.23                       & 1.5      & 9.379652  & 87.59                      & (6.8, 0.6065)      & 5.0      & 0                          \\
4.762602e-27 & 4.558067 & 8.84                       & 3.140998e-29    & 5.137739 & 2.75                       & 2.0      & 11.714818 & 134.30                     & (8.4, 0.6308)      & 5.0      & 0                          \\
4.870843e-27 & 4.633542 & 7.33                       & 3.212384e-29    & 5.114226 & 2.28                       & 2.5      & 13.579730 & 171.59                     & (10.0, 0.6473)     & 5.0      & 0                          \\
4.979084e-27 & 4.708275 & 5.83                       & 3.283771e-29    & 5.090939 & 1.82                       & 3.0      & 14.975248 & 199.50                     & (11.6, 0.6593)     & 5.0      & 0                          \\
5.087325e-27 & 4.782276 & 4.35                       & 3.355157e-29    & 5.067877 & 1.36                       & 3.5      & 16.005400 & 220.11                     & (13.2, 0.6683)     & 5.0      & 0                          \\
5.195566e-27 & 4.855556 & 2.89                       & 3.426543e-29    & 5.045035 & 0.90                       & 4.0      & 16.770707 & 235.41                     & (14.8, 0.6754)     & 5.0      & 0                          \\
5.303807e-27 & 4.928128 & 1.44                       & 3.497930e-29    & 5.022410 & 0.45                       & 4.5      & 17.347245 & 246.94                     & (16.4, 0.6811)     & 5.0      & 0                          \\
5.412048e-27 & 5.000000 & 0.00                       & 3.569316e-29    & 5.000000 & 0.00                       & 5.0      & 17.788712 & 255.77                     & (18.0, 0.6858)     & 5.0      & 0                          \\
5.520289e-27 & 5.071184 & 1.42                       & 3.640702e-29    & 4.977801 & 0.44                       & 5.5      & 18.132347 & 262.65                     & (19.6, 0.6897)     & 5.0      & 0                          \\
5.628530e-27 & 5.141690 & 2.83                       & 3.712089e-29    & 4.955810 & 0.88                       & 6.0      & 18.404035 & 268.08                     & (21.2, 0.6930)     & 5.0      & 0                          \\
5.736771e-27 & 5.211528 & 4.23                       & 3.783475e-29    & 4.934025 & 1.32                       & 6.5      & 18.621957 & 272.44                     & (22.8, 0.6958)     & 5.0      & 0                          \\
5.845012e-27 & 5.280707 & 5.61                       & 3.854861e-29    & 4.912442 & 1.75                       & 7.0      & 18.799073 & 275.98                     & (24.4, 0.6983)     & 5.0      & 0                          \\
5.953253e-27 & 5.349238 & 6.98                       & 3.926248e-29    & 4.891059 & 2.18                       & 7.5      & 18.944755 & 278.90                     & (26.0, 0.7005)     & 5.0      & 0                          \\
6.061493e-27 & 5.417128 & 8.34                       & 3.997634e-29    & 4.869873 & 2.60                       & 8.0      & 19.065890 & 281.32                     & (27.6, 0.7024)     & 5.0      & 0                          \\
6.169734e-27 & 5.484389 & 9.69                       & 4.069020e-29    & 4.848881 & 3.02                       & 8.5      & 19.167610 & 283.35                     & (29.2, 0.7041)     & 5.0      & 0                          \\
6.277975e-27 & 5.551028 & 11.02                      & 4.140407e-29    & 4.828080 & 3.44                       & 9.0      & 19.253795 & 285.08                     & (30.8, 0.7056)     & 5.0      & 0                          \\
6.386216e-27 & 5.617054 & 12.34                      & 4.211793e-29    & 4.807468 & 3.85                       & 9.5      & 19.327415 & 286.55                     & (32.4, 0.7069)     & 5.0      & 0                          \\
6.494457e-27 & 5.682475 & 13.65                      & 4.283179e-29    & 4.787043 & 4.26                       & 10.0     & 19.390770 & 287.82                     & (34.0, 0.7082)     & 5.0      & 0     \\\bottomrule                    
\end{tabular}
\caption{Sensitivity analysis due to the variation of the set of the parameters $(\Delta\mu, E_s, \delta, \phi_d, \sigma)$.}\label{table:1}
\end{table}

\begin{table}
\centering
\footnotesize   
\begin{tabular}{@{}ccccc@{}}
\toprule
Parameter   & Value                   & Unit              & CI (95\%)                        & Method                                                                                                                                                                          \\ \midrule
$E_s$       & 5.412048e-27            & J                 & {[}4.554444e-27, 6.269652e-27{]} & \begin{tabular}[c]{@{}c@{}}Fitted Between $(r_1 = 20 \text{ nm, } \mathcal{B}_1 = 0)$ \\ and $(r_2 = 5 \text{ nm, } \mathcal{B}_1 = 250\text{mT})$ \\ using Eq. 7a\end{tabular} \\ \midrule
$\Delta\mu$ & 3.569316e-29            & J                 & {[}3.000216e-29, 4.138416e-29{]} & \begin{tabular}[c]{@{}c@{}}Fitted Between $(r_1 = 20 \text{ nm, } \mathcal{B}_1 = 0)$\\ and $(r_2 = 5 \text{ nm, } \mathcal{B}_1 = 250\text{mT})$\\ using Eq. 7b\end{tabular}   \\ \midrule
$\sigma$    & 1.2                     & angle             & Not Applicable                   & \begin{tabular}[c]{@{}c@{}}Best match between deterministic ($\sigma = 0$) \\ and experimental broadening (Fig. 3c)\end{tabular}                                                \\ \midrule
$\delta$    & 4                       & a (atomic radius) & Not Applicable                   & Chosen to represent two atomic layers, verified by Packomania data                                                                                                              \\ \midrule
$\phi_b$     & $\frac{\pi}{3\sqrt{2}}$ & dimensionless          & Not Applicable                   & Ideal close packing of spheres.                                                                                                                                                 \\ \midrule
$\phi_d$    & 0.517                   & dimensionless          & Not Applicable                   & Fitted from $1,045$ Packomania data points (Fig. 2b)                                                                                                                             \\ \bottomrule
\end{tabular}
\caption{{A summary of the best fit parameters used to obtain Fig 3 in the main text}}
\end{table}
\newpage

{The Confidence Intervals (CIs) are calculated using the
experimental radii
$r_1 = 20\,\mathrm{nm}$ at $B=0$ and $r_2 = 5\,\mathrm{nm}$ at $B=250\,\mathrm{mT}$.
Each radius is treated as uniformly distributed over
$[\,r \pm 0.5\,\mathrm{nm}\,]$, since the histogram bin width is $1\,\mathrm{nm}$ and no further dispersion
information is provided in Kthiri et al \cite{kthiri2021novel}.
This corresponds to a Type-B (resolution-limited)
standard uncertainty
\[
u_r = \frac{0.5\,\mathrm{nm}}{\sqrt{3}}
= \frac{1}{\sqrt{12}}\,\mathrm{nm}.
\]
Radii are converted to the model coordinate $x = r/a$, and uncertainties are
propagated analytically to the fitted parameters $E_s$ and $\Delta\mu$
using the first-order uncertainty propagation (delta method):
\[
\mathrm{Var}(E_s)
  \approx
  \left(\frac{\partial E_s}{\partial x_1}\right)^2 \mathrm{Var}(x_1)
 + \left(\frac{\partial E_s}{\partial x_2}\right)^2 \mathrm{Var}(x_2)
\]
with an analogous expression for $\Delta\mu$.
Standard errors are $\mathrm{SE}=\sqrt{\mathrm{Var}}$, and
the {95\% CIs} are reported as
\[
\text{CI}_{95\%} = \text{best fit} \pm 1.96\,\mathrm{SE}.
\]
}
\newpage
\bibliography{ref}

\end{document}